\documentclass[showpacs]{iopart}
\usepackage{graphicx,amssymb,amsfonts,latexsym,units,bm}

\begin{document}
\title[Lessons from an exactly solvable Many-Body model]{Symmetry breaking in self-consistent models: Lessons from an exactly solvable many-fermion model}
\author{Emil Prodan}
\address{Department of Physics, Yeshiva University, New York, NY 10016}

\begin{abstract}
This work presents a many-fermion Hamiltonian with the following properties: 1) is exactly solvable, 2) has a second order insulator-metal quantum phase transition, 3) has a well defined mean field approximation and 4) its mean-field ground state displays a liquid-solid transition. The phenomenon of symmetry breaking in fermionic self-consistent models is discussed in the light of these remarkable properties of the many-body model.  
\end{abstract}
\maketitle
\section{Introduction}

Spontaneous translational symmetry breaking, i.e., crystallization, is one of the most interesting problems in condensed matter.  But as stressed by Robert Laughlin in his book \cite{Laughlin:2005fu}, the liquid-solid transition remains one of the outstanding unsolved problems in theoretical physics. To date, there is no quantum mechanical model in which crystallization can be observed first hand.

In contradistinction, translational symmetry breaking occurs quite often in the self-consistent models of the condensed matter. For repulsive interactions, Ref.~\cite{Bach:1994kk} showed that the energy levels in the unrestricted Hartree-Fock approximation are always fully populated. This result automatically implies symmetry breaking whenever the last occupied level of the translational invariant Hartree-Fock Hamiltonian is only partially populated. Ref.~\cite{Prodan:2001gv} proved the existence of symmetry breaking in the fermionic Hartree approximation for short, attractive interactions. The precise conditions and the mechanism of symmetry breaking in the Kohn-Sham equations was discussed in Ref.~\cite{Prodan:2005qg}. The Wigner crystallization of the electron liquid was studied in the Refs.~\cite{Likos:1997ud} through a combination Monte-Carlo and Density Functional Theory calculations. The symmetry breaking in the mean field approximation of the bossonic Hubbard model was recently discussed in Ref.~\cite{Oelkers:2007lr}. The modern theory of freezing \cite{Ramakrishnan:1997lh,Baus:1990wq} is based on the assumption that the linear density response equation of the liquid displays a crystalized self-consistent solution as soon as the system crosses the liquid-solid phase boundary. And the list can continue.

The crystallization in fermionic self-consistent models, such as Hartree, Hartree-Fock and Kohn-Sham, is triggered by a strong coupling between the bare electrons and holes at the opposite sites of the Fermi surface. In well defined conditions, these models were shown to display robust fixed points that break the translational symmetry of the many-body Hamiltonian \cite{Prodan:2001gv,Prodan:2005qg}. Almost like a rule, for metallic systems, the crystallization was observed to be accompanied by a gap opening in the mean-field spectrum.  

In this work, we deal with the translational symmetry breaking in self-interacting 1 dimensional (1D) quantum liquids. According to a result by Mermin \cite{Mermin:1968bv}, true crystallization cannot occur in 1 and 2D for specific short range interacting systems. This result, however, does not entirely exclude crystallization in 1D, as shown by explicit 1D models \cite{Haldane:1980cr}, or mathematically rigorous statements \cite{Giuliani:2006dq,Giuliani:2007rr,Giuliani:2009lq}. In any case, the issue is extremely puzzling because the self-consistent models have a tendency to display robust crystallization precisely in lower dimensions.

Here we propose a many-body Hamiltonian, which is a band insulator with a particular two-body interaction that takes into account only the electron-hole coupling mentioned above. While the model is fairly simple, it still displays interesting features such as a second order singularity of the ground state energy as function of interaction strength, an insulator-metal phase transition and also crystallization in its mean-field approximation. Based on these features, we can explicitly see how how the crystallization seen in the mean-fied approximation relates to the exact many-body solution.

\section{The many-body model} 

We start with several considerations that will allow us to place the many-body model introduced here relative to the Luttinger liquid \cite{Haldane:1981wd}. The Luttinger liquid concept is now generally accepted to apply to all conducting spinless fermion systems in one dimension \cite{Haldane:1981wd}, whenever such conducting behavior can be established. The Luttinger liquid state is, however, known to be unstable against an insulating pinned charge-density-wave state \cite{Haldane:1981wd} and in fact the full phase diagram of the 1D Fermi systems is not presently known. It will be interesting to establish a solvable many-body model that is relevant to a region of the phase diagram not covered by the Luttinger liquid concept. We will try to argue that this is the case for the present model.

Let us start from the 1D spinless Fermi gas on a circle of length $L$,
self-interacting via a two-body potential $v$. The dynamics of the gas is generated 
by the following general Hamiltonian:
\begin{equation}\label{Hamiltonian}
    H_{\mbox{\tiny{Gen}}}=\frac{1}{2}\sum_{k} (k^2-k_{F}^2)a^{\dagger}_{k}a_{k}
    +\frac{1}{L}\sum_{k}v_k\hat{\rho}_{k}\hat{\rho}_{-k} \ (\equiv T+V),
\end{equation}
where $a^{\dagger}_{k}$ creats a fermion in the state
$\phi_{k}(x)=e^{ikx}/\sqrt{L}$, $k=2n\pi/L$, $n=0,\pm 1,\ldots$, and
$\hat{\rho}_{k}=\sum_{k^\prime}a^{\dagger}_{k^\prime+k}a_{k^\prime}$. 
The Fermi wave-vector, $k_F$, is assumed equal to one of the $k_n$ vectors. 
We follow Ref.~\cite{Heidenreieh:1980oq} and introduce the Luttinger variables as:
\begin{eqnarray}
    a_{1q}\equiv  a_{q+k_{F}},\ q\in [-k_{F},\infty], \ 
    a_{2q}\equiv  a_{q-k_{F}},\ q\in [-\infty,k_{F}).
\end{eqnarray}
Note that no fictitious states have been introduced. With these new
variables, the kinetic energy becomes
\begin{equation}
    T=\sum_{q\geq-k_F} q \left(k_F+q/2\right)a^{\dagger}_{1q}a_{1q}
    -\sum_{q<k_F} q \left(k_F-q/2\right)a^{\dagger}_{2q}a_{2q}
\end{equation}
and
\begin{eqnarray}\label{rho}
    \hat{\rho}_{k}&=&\sum_{q=-k_{F}}^{\infty}a^{\dagger}_{1q+k}a_{1q}
    +\sum_{q=-\infty}^{k_{F}-k}a^{\dagger}_{2q+k}a_{2q}
    +\sum_{q=-k}^{0}a^{\dagger}_{1q+k-k_{F}}a_{2q+k_{F}}.
\end{eqnarray}
Our observation is that, in the limit $k_{F}$$\rightarrow$$\infty$, the third therm in Eq.~(\ref{rho}) 
commutes with all $a_{1q}$ and $a_{2q}$. Mathematically, this is
equivalent to say that it goes weakly to zero. Consequently,
$\hat{\rho}_{k}\rightarrow
    \hat{\rho}_{1k}+\hat{\rho}_{2k}$ in the limit $k_F\rightarrow
    \infty$, with
\begin{equation}
    \hat{\rho}_{1k}\equiv\sum_{q=-\infty}^{\infty}a^{\dagger}_{1q+k}a_{1q}, \ \hat{\rho}_{2k}\equiv\sum_{q=-\infty}^{\infty}a^{\dagger}_{2q+k}a_{2q},
\end{equation}
being the classic density operators appearing in the Luttinger model. We can immediately conclude that the Luttinger
model \cite{Mattis:1965eu},
\begin{eqnarray}\label{Luttinger}
H_{\cal L}=\sum_q q(
    a^{\dagger}_{1q}a_{1q}-a^{\dagger}_{2q}a_{2q})
    +\frac{1}{L}\sum_k v_k(
    \hat{\rho}_{1k}+\hat{\rho}_{2k})(
    \hat{\rho}_{1-k}+\hat{\rho}_{2-k}),
\end{eqnarray}
can be viewed as the following limit of the 1D fermi gas:
\begin{equation}
   H_{\cal L}=\lim_{k_F\rightarrow \infty}
   \frac{1}{k_F}\left[T+k_{F}V\right].
\end{equation}
According to this observation, the Luttinger liquid concept is for sure relevant in the limit of large particle densities (large $k_F$). But in this limit, the Fourier component $v_{2k_{F}}$ of any well behaved interaction goes to zero, and in fact the exactly solvable Luttinger
model excludes the coupling between the states near $\pm k_F$, contained in the third term of Eq.~(\ref{rho}). This coupling is precisely at the origin of the charge-density-wave instability.

Our many-body model involves the extreme case of a two-body potential,
\begin{equation}
    v_k=v_{0}[\delta_{k,2k_{F}}+\delta_{k,-2k_{F}}],
\end{equation}
which zooms into the coupling of the $2k_{F}$ spaced one particle states and ignores anything else. Only the case $v_0$$<$$0$ will be examined in this work, which is interesting for the symmetry breaking problem. We also restrict the one particle Hilbert states to $k$'s within the intervals $[-2k_{F},2k_{F})$. Therefore, we ignore in the original Hamiltonian Eq.~(\ref{Hamiltonian}) all 
$a_{k}$ and $a^{\dagger}_{k}$ with index outside the interval $[-2k_{F},2k_{F})$.  With the notation
$\hat{\rho}^{\pm}_k\equiv \frac{1}{2}[\hat{\rho}_{k}\pm
\hat{\rho}_{-k}]$, the interaction potential of Eq.~\ref{Hamiltonian} becomes
\begin{equation}
    V\rightarrow\frac{v_{0}}{L}[(\hat{\rho}^{(+)}_{2k_F})^2-(\hat{\rho}^{(-)}_{2k_F})^2].
\end{equation}
Given our constraints on the $k$ wavenumber, the operators $\hat{\rho}^{(\pm)}_{2k_{F}}$ involve only the third term of $\hat{\rho}_k$ in the expansion of Eq.~(\ref{rho}), i.e., exactly the term that is neglected in the exactly solvable Luttinger model.

Before we assemble the total Hamiltonian, we need to discuss the one-particle Hamiltonian: 
\begin{equation}
    H_{0}= \sum_{-2k_{F}\leq
    k<2k_{F}}(\varepsilon_{k}-\varepsilon_{F})a^{\dagger}_{k}a_{k}.
\end{equation}
We would like to start from an insulating state, i.e., exactly opposite to the Luttinger liquid regime. We propose the following dispersion:
\begin{equation}
\varepsilon_{k}=\left \{
\begin{array}{c}
\varepsilon_{F}-\Delta \ \mbox{for} \  |k|<k_{F} \smallskip \\
\varepsilon_{F}+\Delta \ \mbox{for} \ |k|>k_{F},
\end{array} \right.
\end{equation}
in which case $H_0$ describes an insulator with two non-dispersive bands, viewed in an extended Brillouin zone, separated by a gap $2\Delta$. The proposed full many-body Hamiltonian is
\begin{equation}
H= \sum_{-2k_{F}\leq
    k<2k_{F}}(\varepsilon_{k}-\varepsilon_{F})a^{\dagger}_{k}a_{k}+\frac{v_{0}}{L}[(\hat{\rho}^{(+)}_{2k_F})^2-(\hat{\rho}^{(-)}_{2k_F})^2].
\end{equation}
To summarize, the above Hamiltonian assumes an energy cut-off at $2k_F$, or a two band approximation, non-dispersive bands separated by a gap $2\Delta$, and a singular two-body interaction that takes into account only the electron-hole couplings that were argued to be relevant for the translational symmetry breaking.

\section{Diagonalizing the Hamiltonian}

It is convenient to render $k$ from $\pm k_F$ and express $a_{k}$ in terms of creation and
destruction operators with respect to the ground state $\Psi_0$ of $H_0$, for which all the states with
$k\in[-k_{F},k_{F})$ are occupied:
\begin{equation}
    a_{q-k_{F}}=\left\{
    \begin{array}{l}
        b_{q},\ -k_{F}\leq q<0 \medskip \\
        c^{\dagger}_{q},\ 0\leq q<k_{F}
    \end{array}
    \right.
, \ 
    a_{q+k_{F}}=\left\{
    \begin{array}{l}
        c^{\dagger}_{q},\ -k_{F}\leq q<0 \medskip \\
        b_{q},\ 0\leq q<k_{F}.
    \end{array}
    \right.
    \label{new2}
\end{equation}
In terms of the new creation and destruction operators,
\begin{equation}
    \hat{\rho}^{(+)}_{2k_F}=\frac{1}{2}\sum_{q}[b^{\dagger}_{q}c^{\dagger}_{q}+c_{q}b_{q}], \ 
    \hat{\rho}^{(-)}_{2k_F}=\frac{1}{2}\sum_{
    q} \chi(q)[b^{\dagger}_{q}c^{\dagger}_{q}-c_{q}b_{q}],
\end{equation}
where $-k_{F}\leq
    q<k_{F}$ and $\chi(q)=-1$ for $-k_{F}\leq q<0$ and 1 for $0\leq q<k_{F}$. The interaction potential can be explicitly diagonalized by considering
\begin{eqnarray}\label{TheL}
    L_{1}=\hat{\rho}^{(+)}_{2k_F}, \ \ L_{2}=-i\hat{\rho}^{(-)}_{2k_F}, \ 
    L_{3}=\frac{1}{2}\sum\limits_{-k_{F}\leq
    q<k_{F}}\chi(q)[b^{\dagger}_{q}b_{q}-c_{q}c^{\dagger}_{q}].
\end{eqnarray}
A simple algebra shows that $[L_{i},L_{j}]=i\varepsilon_{ijk}L_{k}$
and that
\begin{equation}
    V=\frac{v_0}{L}[\vec{L}^2-L_3^2].
\end{equation}
The eigenstates of $V$ are the usual $|JM\rangle$ states, where the highest weight states $|JJ\rangle$ are in general not unique, and one can show 
that $J_{\max}=\frac{1}{2}N_0$, with $N_0$ the number of
particles. $N_0$ is an even number because of our careful choice $-2k_F\leq k < 2k_F$. The following eigenvectors can be computed explicitly:
\begin{equation}
    |J_{\max},J_{\max}\rangle = \prod_{0\leq q<k_F}b_q^\dagger
    c_q^\dagger\Psi_0, \ 
    \ |J_{\max}, -J_{\max}\rangle = \prod_{-k_F\leq q<0}b_q^\dagger
    c_q^\dagger\Psi_0,
\end{equation}
and the manifold $|J_{\max}M\rangle$ can be shown to be non-degenerate (i.e. there is just one highest weight vector $|J_{\max}J_{\max}\rangle$. The lowest energy state of $V$ is $|J_{\max}0\rangle$, which can be obtained from the above vectors by applying the $L_\mp$ operators
$\frac{1}{2}N_0$ times.

We now consider the whole Hamiltonian $H$=$H_0$+$V$. The new representation of $H_0$ is
\begin{equation}
H_{0}\rightarrow\Delta\sum_q[b^{\dagger}_{q}b_{q}-c_{q}c^{\dagger}_{q}].
\end{equation}
$H_{0}$ commutes with $L_3$ but not with $\vec{L}^2$. It is interesting to notice that there is a
competition between $V$, which favors electron-hole pair formations, and $H_0$, which does not. As we shall see, this competition will ultimately lead to a quantum phase transition. The model can be solved by employing the SO(4) Lie algebra. Indeed, if we define
\begin{equation}
\begin{array}{l}
    K_1=\frac{1}{2}\sum_{-k_{F}\leq
    q<k_{F}}\chi(q)[b^{\dagger}_{q}c^{\dagger}_{q}+c_{q}b_{q}] \medskip \\
    K_2=-\frac{i}{2}\sum_{-k_{F}\leq
    q<k_{F}}[b^{\dagger}_{q}c^{\dagger}_{q}-c_{q}b_{q}] \medskip \\
    K_3=\frac{1}{2}\sum_{-k_{F}\leq
    q<k_{F}}[b^{\dagger}_{q}b_{q}-c_{q}c^{\dagger}_{q}],
\end{array}
\end{equation}
then
\begin{eqnarray}
    [ L_i, L_j ]=
    [ K_i, K_j ]=i\epsilon_{ijk}L_k, \
    [ L_i,K_j ]=i\epsilon_{ijk}K_k,
\end{eqnarray}
precisely the SO(4) Lie algebra. A straightforward calculation gives
\begin{equation}
    H=2\Delta \hat{K}_3+\frac{v_{0}}{L}[\hat{L}^2-\hat{L}_{3}^2].
\end{equation}
If $\vec{X}\equiv\frac{1}{2}(\vec{L}+\vec{K})$ and
$\vec{Y}\equiv\frac{1}{2}(\vec{L}-\vec{K})$, then
\begin{equation}
[X_i,X_j]=i\epsilon_{ijk}X_k, \ \ [Y_i,Y_j]=i\epsilon_{ijk}Y_k, \ \ [X_i,Y_j]=0
\end{equation}
and 
\begin{equation}
    [H,\vec{X}^2]=[H,\vec{Y}^2]=[H,X_3+Y_3]=0.
\end{equation}
The operators $\vec{X}$ and $\vec{Y}$ commute with the particle number operator, therefore we can restrict the discussion to the quantum states with precisely $N_0$ particles.

Let us denote the common eigenstates of $\vec{X}^2$ and $X_3$ by $|jm\rangle$ and the common eigenstates of $\vec{Y}^2$ and $Y_3$ by 
$|j^\prime m^\prime\rangle$. We verified that the maximum values of $j$ and $j'$ are $j_{\max}$=$j^\prime_{\max}$=$\frac{1}{4}N_0$.
Now, for fixed $j$, $j^\prime$ and $M$, the vector space 
spanned by
$\phi_m^{(jj'M)}=|jm\rangle\otimes|j'M-m\rangle$ with $m$ taking all allowed values, is invariant for $H$:
\begin{equation}\label{Hreduced}
    H\phi_m=\frac{v_0}{L}(a_{m-\frac{1}{2}}\phi_{m-1}-2a_m\phi_m+a_{m+\frac{1}{2}}\phi_{m+1})+v_m   \phi_m,
\end{equation}
where, we dropped the upper indices to ease the notation. The coefficients appearing above are given by:
\begin{equation}\label{descrete}
\begin{array}{l}
    a_m=\sqrt{[(j+1/2)^2-m^2][(j^\prime+1/2)^2-(m-M)^2]} \medskip \\
    v_m=\frac{v_0}{L}[j(j+1)+j^\prime(j^\prime+1)-M^2+2m(M-m)] \medskip \\
    \ \ \ \ \ \ +2\Delta(2m-M)+2\frac{v_0}{L}a_m.
\end{array}
\end{equation}
The vector $\phi_m$ must be set to zero if $|m|>j$ or $|M-m|>j'$, a statement that clarifies the allowed values of $m$.
Eq.~(\ref{Hreduced}) allows one to calculated the whole energy spectrum of $H$ and solving it is no more complicated than diaganolizing a one-particle tight-binding model in 1D. The many-body eigenvalues fall into distinct manifolds that can be labeled by $j$, $j'$ and $M$. To 
calculate thermodynamic functions, we also need to calculate the degeneracy of each manifold and for that we need to compute how many highest weight vectors $|j j\rangle \otimes |j^\prime j^\prime\rangle$ are there for each $j$ and $j'$. This will not be done here.

\section{Thermodynamic limit}

Things greatly simplify in the thermodynamic limit $L$$\rightarrow$$\infty$, when 
$j$, $j^\prime$ and $M$ take macroscopic values (proportional to $L$). For 
large $L$, we normalize $m$ by $j$ and work with $x\equiv m/j$ ($|x|\leq 1$) as our variable, which now becomes continuous. For fixed $j$, $j'$ and $M$ and with the representation $\Psi=\sum_m c_m \phi_m$, where $c$ becomes a function of $x$, the action of the Hamiltonian per unit length becomes:
\begin{equation}\label{Hami}
    \frac{1}{L}Hc(x)=\frac{v_0}{L^2}\partial_x a(x) \partial_x
    c(x)+v(x)c(x).
\end{equation}
For each manifold $\{j,j^\prime,M\}$, the potentials $a(x)$ and
$v(x)$ can be easily derived from Eq. (\ref{descrete}),
\begin{eqnarray}
    a(x)&=&\sqrt{(1-x^2)[(j^\prime/j)^2-(x-M/j)^2]} \nonumber \\
    v(x)&=&v_0\left(j/L\right)^2[1+(j^\prime/j)^2-(x-M/j)^2-x^2] \nonumber \\
    &+&2\Delta (j/L)[2x-M/j]+2v_0\left(j/L\right)^2a(x).
\end{eqnarray}
Eq.~(\ref{Hami}) is defined on the interval where the factor under the 
square root in $a(x)$ is positive and
zero boundary must be imposed at the ends of this interval.

From Eq.~(\ref{Hami}), one can see that, when $L$$\rightarrow$$\infty$, the energy per unit length is given by $v(x)$. The ground state is contained in the manifold
$j=j^\prime=j_{\max}=\frac{1}{4}N_0$, and $M=0$. This can be established analytically but we have also verified the statement numerically. For this manifold, $v(x)$ is
equal to:
\begin{equation}
    v_{\min}(x)=n_0^2v_0(1-x^2)/4+n_0\Delta x,
\end{equation}
with $n_0\equiv \lim\limits_{L\rightarrow \infty}N_0/L=k_F/\pi$. Given that $x$ is 
constrained to $|x|\leq 1$, the minimum of this potential, which defines the ground state energy
per unit length, corresponds to 
$x_0=\max\{\frac{2\Delta}{n_0v_0},-1\}$
and the energy per unit length is:
\begin{equation}\label{GR}
	\lim\limits_{L\rightarrow\infty}\frac{E_0}{L}=
\left\{\begin{array}{l}
	 -n_0\Delta, \ 2\Delta>n_0|v_0| \medskip \\
	n_0 \Delta[\frac{n_0v_0}{4\Delta}+\frac{\Delta}{n_0v_0}], \ 2\Delta<n_0|v_0|.
\end{array}
 \right.
\end{equation}
Thus, in the thermodynamic limit, there is a sharp change in the behavior of the ground 
state energy as a function of the model's paramaters. We have verified numerically that, indeed, $E_0/L$ converges to the values predicted in Eq.~\ref{GR}. 

To get more insight, we calculated  the expectation value of $H_0$ ($\equiv 2\Delta K_3$) on the ground state of $H$, divided by $L$:
\begin{equation}
	\lim\limits_{L\rightarrow \infty} \frac{\langle H_0 \rangle}{L}=
	\left\{\begin{array}{l}
	-n_0 \Delta, \ 2\Delta>n_0|v_0| \medskip \\
	2\Delta^2/v_0, \ 2\Delta < n_0|v_0|.
	\end{array}\right.
\end{equation}
From here, we conclude that for $v_0$ larger than the critical value $v_c=2\Delta/n_0$, we have 
macroscopic occupation of states with $|k|>k_F$.

\section{Mean field analysis} 

We define the mean field problem by the following 
substitution:
\begin{equation}
	V\rightarrow \frac{v_0}{L}[\langle\hat{\rho}^{(+)}_{2k_F}\rangle 
	\hat{\rho}^{(+)}_{2k_F}-\langle\hat{\rho}^{(-)}_{2k_F}\rangle \hat{\rho}^{(-)}_{2k_F}],
\end{equation}
where $\langle\hat{\rho}^{(\pm)}_{2k_F}\rangle$ denotes the expectation value of 
$\hat{\rho}^{(\pm)}_{2k_F}$ on the ground state of the mean field Hamiltonian, which 
has to be calculated self-consistently. If $\alpha$$\equiv$$\langle\hat{\rho}_{2k_F}\rangle$, then $\langle \hat{\rho}^{(+)}_{2k_F}\rangle$=$\mbox{Re}[\alpha]$
and $\langle\hat{\rho}^{(-)}_{2k_F}\rangle$=$i\mbox{Im}[\alpha]$. We assume in the 
following that $\alpha$ is a real positive number. The mean field Hamiltonian becomes:
\begin{equation}
	H_{\mbox{\tiny{MF}}}=\sum\limits_q \left\{ \Delta (b_q^\dagger 
	b_q-c_qc_q^\dagger)+\frac{\alpha v_0}{L}(b_q^\dagger 
	c_q^\dagger+c_qb_q)\right\}.
\end{equation}
We can diagonalize the quadratic mean field Hamiltonian by using the following 
Bogoliubov substitution:
\begin{equation}
	b_q=\cos \theta \tilde{b}_q+\sin \theta \tilde{c}_q^\dagger, \ 
	c_q=-\sin \theta \tilde{b}_q^\dagger+\cos \theta \tilde{c}_q,
\end{equation}
with
\begin{equation}\label{alpha}
	\tan 2 \theta=-\frac{\alpha v_0}{\Delta L}, \ \ \theta\in [0,\pi/2].
\end{equation}
The mean field Hamiltonian becomes
\begin{equation}
	H_{\mbox{\tiny{MF}}}=\frac{\Delta}{\cos 
	2\theta}\sum\limits_q(\tilde{b}_q^\dagger\tilde{b}_q+
	\tilde{c}_q^\dagger\tilde{c}_q-1),
\end{equation}
and
\begin{equation}
	\hat{\rho}^+_{2k_F}=-\frac{\sin
        2\theta}{2}\sum\limits_q(\tilde{b}_q^\dagger\tilde{b}_q-
        \tilde{c}_q\tilde{c}_q^\dagger) 
	+\frac{\cos2\theta}{2}\sum\limits_q
	(\tilde{b}_q^\dagger\tilde{c}_q^\dagger+
        \tilde{c}_q\tilde{b}_q).\nonumber
\end{equation}
It remains to determine $\alpha\equiv 
\langle\hat{\rho}_{2k_F}\rangle$, self-consitently. Of course, there is always the trivial solution $\alpha$=0 but we will show that, if $v_0$ is negative enough, the mean field approximation has nontrivial self-consistent solutions.

Indeed, let us assume $1\geq\cos 2\theta>0$, in which case the ground state of $H_{\mbox{\tiny{MF}}}$ has all the $\tilde{b}_q$ and 
$\tilde{c}_q$ states empty and the expectation value of $\hat{\rho}^+_{2k_F}$ 
becomes
\begin{equation}
	\alpha=\nicefrac{1}{2} N_0\sin 2\theta,
\end{equation}
which together with Eq.~(\ref{alpha}) leads to
\begin{equation}
	\cos 2\theta=-\frac{2\Delta}{n_0v_0} \ \mbox{and} \  \alpha= \frac{N_0}{2} \sqrt{1-\left ( \frac{2\Delta}{n_0v_0} \right )^2}.
\end{equation}
This self-consistent solution is in line with the starting assumptions that $\alpha>0$ and $\cos 2\theta>0$, therefore it is a valid solution. This solution exists as long as 
$2\Delta<n_0|v_0|$, otherwise the the cosine will exceed its maximum value of 1. 

The mean field approximation predicts a ground state energy per unit 
length
\begin{equation}
        \lim\limits_{L\rightarrow\infty}\frac{E_0^{\mbox{\tiny{MF}}}}{L}=\left\{\begin{array}{l}
         -n_0\Delta, \ 2\Delta>n_0|v_0| \medskip \\
        n_0^2 v_0/2, \ 2\Delta<n_0|v_0|,
\end{array}
 \right.
\end{equation}
which also displays a sharp transition. 

\section{Discussion}

\begin{figure}
\center
  \includegraphics[width=12.5cm]{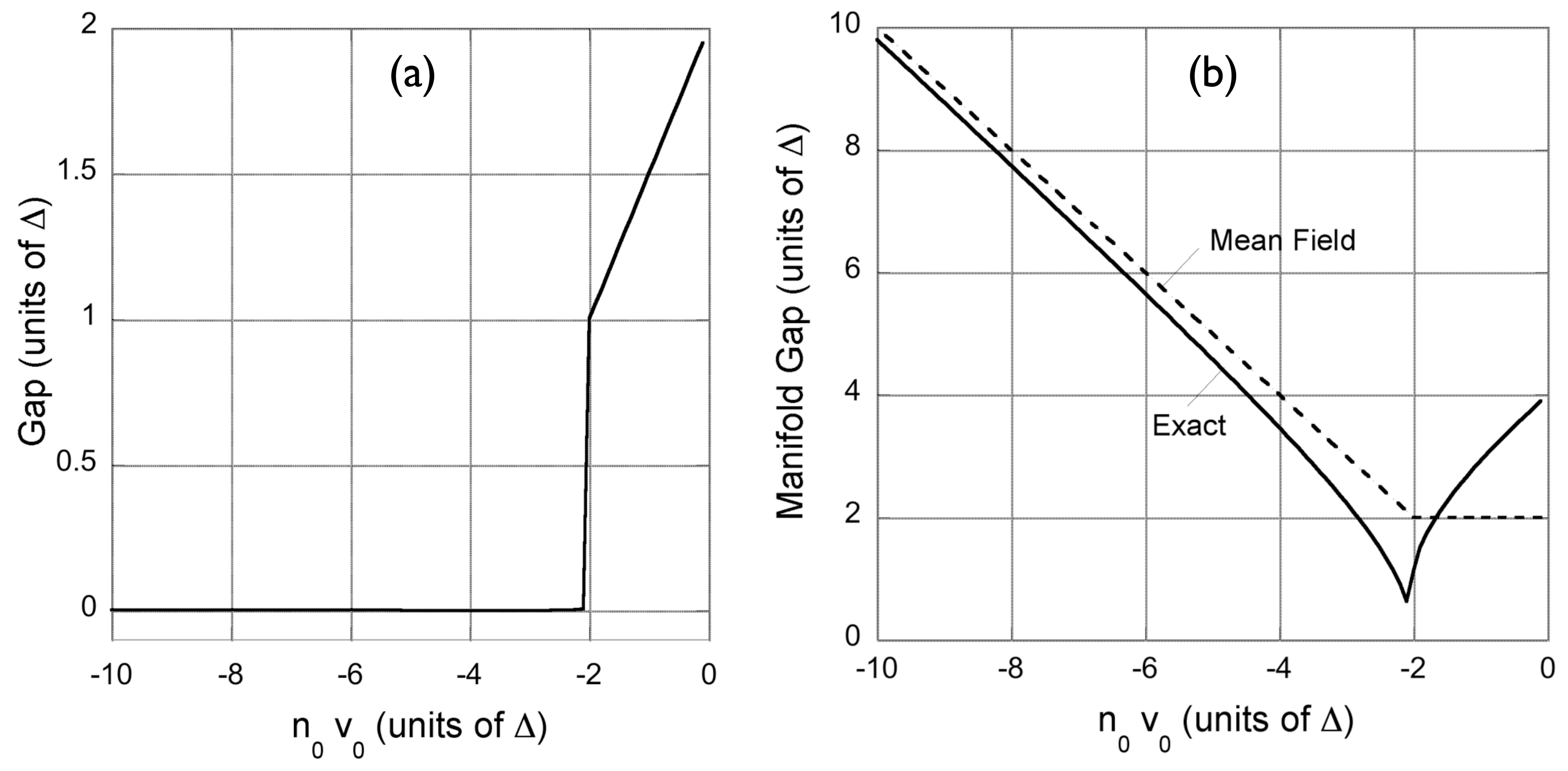}\\
  \caption{(a) The energy gap of the many-body Hamiltonian $H$ as function of $n_0v_0$. (b) The energy gap between the first and second levels of the manifold $\{j_{\max}j_{\max}M=0\}$ (solid line) and the energy gap of the mean field Hamiltonian $H_{\mbox{\tiny{MF}}}$ (dashed line). Both are plotted as functions of $n_0v_0$. The only input for these calculations was $N_0$=4000.}
 \label{Fig1}
\end{figure}

The first observation is that both the exact and the mean field treatments predict the same critical value of $v_0$, where the character of the ground state changes. But there are several differences between the two. The exact ground state energy per unit length $E_0/L$ and its first derivative with respect to $v_0$ are smooth as $v_0$ crosses the critical value. The second derivative, however, has a jump. For the mean field approximation, already the first derivative of $E_0^{\mbox{\tiny{MF}}}/L$ with respect to $v_0$ has a jump at $v_c=-2\Delta/n_0$.

But the most interesting difference concerns the gap of the energy spectrum. In Fig.~\ref{Fig1}(a) we plot the gap of $H$ (no length normalization) as function of $n_0v_0$, with $N_0$ fixed at 4000. The gap of $H$ is given by the difference between the ground state energies of the manifolds $\{j_{\max},j_{\max},M=\pm 1 \}$ and $\{j_{\max},j_{\max},M=0\}$. As one can see, the exact solution indicates a transition to an un-gapped system. In fact, above $v_c$, the ground states of the manifolds  $\{j_{\max},j_{\max},|M|\ge 0 \}$ form a continuous sequence of energy levels. Therefore, at $v_c$ we have a true insulator-metal transition.   

As already mentioned in the Introduction, the non-trivial mean field solutions found so far predict metal-insulator transitions at the symmetry breaking. The situation is similar in the present case, even though we start from a band insulator. In Fig.~\ref{Fig1}(b) we plot the gap of $H_{\mbox{\tiny{MF}}}$ as function of $n_0v_0$. We see the mean field gap first decreasing, but never reaching zero, and then increasing robustly. Therefore, the mean field ground state becomes more insulating as $n_0v_0$ is increased. At first sight, this seems to be in total contradiction with the insulator-metal transition of the exact ground state. But if we look at the gap within one manifold, we actually see a very strong correlation. In Fig.~\ref{Fig1}(b) we also plot the difference between the second and first eigenvalues within the $\{j_{\max},j_{\max},M=0\}$ manifold. As one can see, the mean field gap actually reproduces the manifold gap quite precisely as $n_0v_0$ gets larger.

The non-trivial self-consistent solution of the mean field problem implies crystallization. Indeed, the nontrivial expectation value of $\langle \phi_{\mbox{\tiny{MF}}}|\hat{\rho}(x)|\phi_{\mbox{\tiny{MF}}}\rangle =\frac{1}{L}\sum_k e^{-ikx}\langle\hat{\rho}_k \rangle_{\mbox{\tiny{MF}}}$ leads to a modulated particle density:
\begin{equation}
n(x)=  n_0\left (1+\cos(2k_Fx)\sqrt{1-\left ( \frac{2\Delta}{n_0v_0} \right )^2}\right ).
\end{equation}
It is easy to check that if $\alpha$ is a self-consistent solution, then $e^{i\varphi}\alpha$ is a self-consistent solution too, and the only change in $n(x)$ is the appearance of the phase factor inside the cosine: $\cos(2k_F x)$$\rightarrow$$\cos(2k_F x + \varphi)$. The expectation value of $\hat{\rho}(x)$ on the exact ground state, however, is independent of $x$ and equal to $n_0$. In other words, crystallization is absent in the exact ground state. But as already pointed out, the system becomes gapless so it will be interesting to see if crystallization can occur in the low lying excited states.

To answer the last question, we start from the action of $\hat{\rho}^{(\pm)}_{2k_F}$:
\begin{equation}
\hat{\rho}^{(\pm)}_{2k_F}\phi_m^{jj'M}=A_{M,m}^{(\pm)} \phi_{m\pm1}^{jj'M\pm 2},
\end{equation} 
with $A_{m,M}^{(\pm)}=\sqrt{(j\mp m)(j\pm m+1)}+\sqrt{(j'\mp (M-m))(j'\pm (M-m)+1)}$. In other words, $\hat{\rho}^{(+)}_{2k_F}$ takes the $\{jj'M\}$ manifold into the $\{jj'M+2\}$ manifold and  $\hat{\rho}^{(-)}_{2k_F}$ takes the $\{jj'M\}$ manifold into the $\{jj'M-2\}$ manifold. Given these actions, and the fact that the ground state energies of the manifolds $\{j_{\max}j_{\max}M=\pm1\}$ are the same and converging to the absolute ground state energy in the thermodynamic limit, we reached the conclusion that the mean field ground state actually relates to the states:
\begin{equation}\label{cry}
\frac{1}{\sqrt{2}} \Psi_0^{(j_{\max}j_{\max}M= - 1)}+\frac{e^{i\varphi}}{\sqrt{2}} \Psi_0^{(j_{\max}j_{\max}M=+1)},
\end{equation}
where $\Psi_0^{jj'M}$ denotes the ground state within the manifold $\{jj'M\}$.

\begin{figure}
\center
  \includegraphics[width=12.5cm]{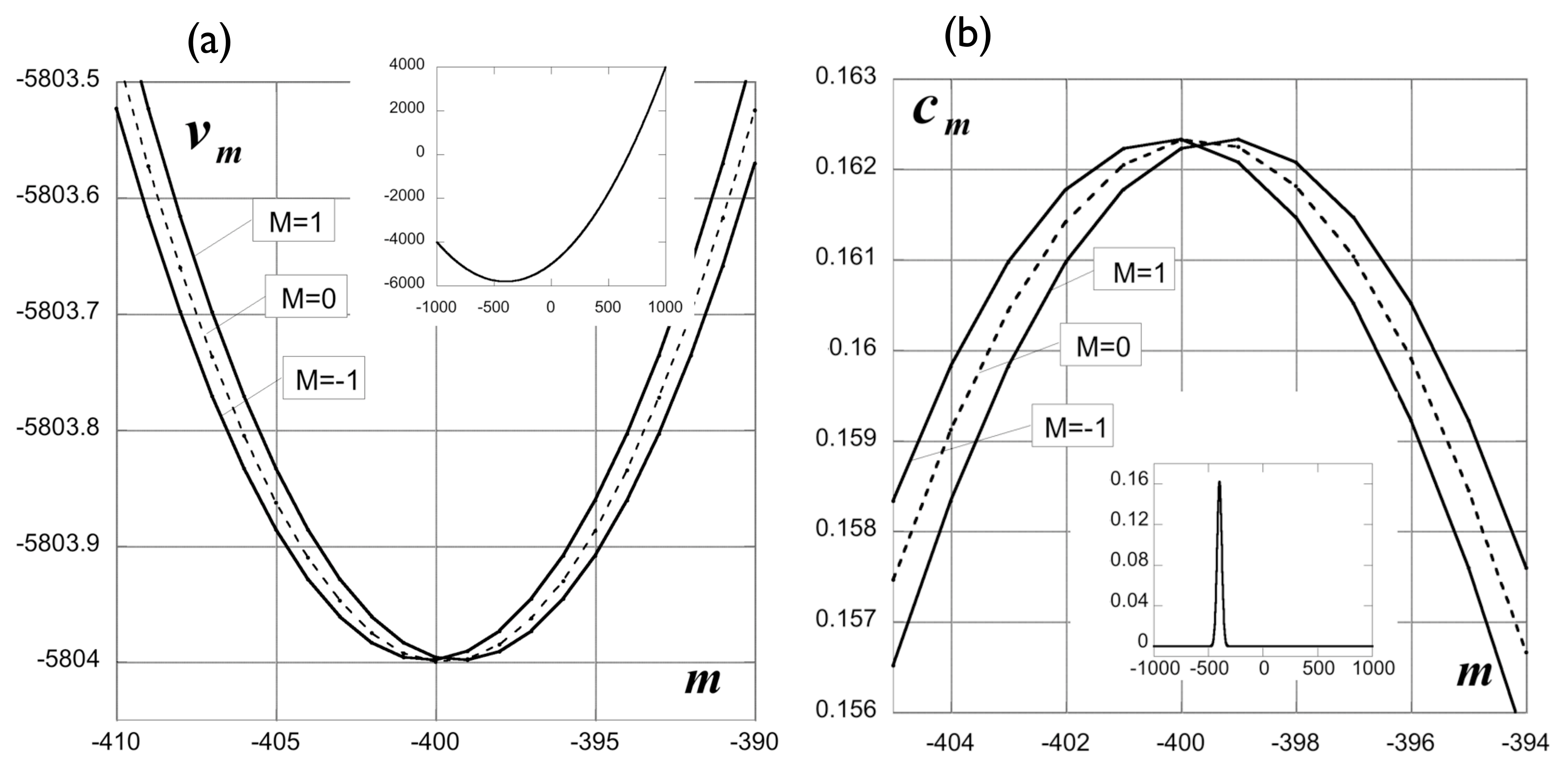}
  \caption{(a) The potential $v_m$ corresponding to the manifolds $j=j'=j_{\max}$  and $M=0,\pm 1$. The inset gives a global view of $v_m$ for $M=0$. (b) The amplitudes $c_m$ of the ground states $\Psi_0$ corresponding to the manifolds $\{(j_{\max} j_{\max} M=0,\pm 1)\}$. The inset gives a global view of $\Psi_0$ for $\{(j_{\max} j_{\max} M=0\}$. The input parameters were $N_0$=4000 and $n_0v_0=-5\Delta$.}.
 \label{Fig2}
\end{figure}

In Fig.~\ref{Fig2} we present plots of the eigenvectors discussed above. Although Eq.~\ref{Hami} suggests that $H$ becomes just a multiplicative potential in the thermodynamics, the eigenvectors actually display a quantum spread even for large $L$. This is because the potential $v_m$ becomes more and more flat as $L$ is taken to infinity. In Fig.~\ref{Fig2}(a) we show the potential $v_m$ for the manifolds $\{j_{\max}j_{\max}M=0,\pm1\}$. The inset shows a global picture of $v_m$ for $n_0v_0=-5\Delta$ and manifold $\{j_{\max}j_{\max}M=0\}$. The main figure is a zoom into the bottom of the potential. One should notice the very small scale on the vertical axis, which shows how flat the potentials are. One should also notice that the minimum of the potentials occurs at certain values of $m$, that shift to right as $M$ is increased. Although the potential of the manifold $\{j_{\max}j_{\max}M=0\}$ is lower, this cannot be distinguished in the figure because the differences are miniscule.

In Fig.~\ref{Fig2}(b) we show the coefficients $c_m$ of the expansion $\Psi_0^{(jj'M)}$=$\sum_m c_m \phi_m^{(jj'M)}$. Again, the inset gives a global picture of the eigenvectors, while the main figure zooms into the region where $c_m$'s take large values. One should notice the considerable spread of the eigenvectors and that the maximum value of $c_m$ occurs at certain $m$'s that drift to the right  by about one unit when $M$ is increased by two units. This drift is essential in order to see large expectation values of $\hat{\rho}^{(\pm)}_{2k_F}$, since these operators increase/decrease $m$ by exactly one unit when applied on $\phi_m^{(jj'M)}$. And indeed, we have verified numerically that the expectation value of $\hat{\rho}_{2k_F}$=$\hat{\rho}^{(+)}_{2k_F}$+$\hat{\rho}^{(-)}_{2k_F}$ on the vector written in Eq.~\ref{cry} is consistent with the value of $\alpha$ provided by the mean field approximation, within 3 significant decimals.

\section{Conclusions}

The exactly solvable 1D many-fermion model have given us unprecedented insight into the phenomenon of translational symmetry breaking observed in mean field approximations. The mean field of the present model has a robust solution displaying the crystallization of the 1D fermions, provided the strength of the interaction exceeds a certain threshold. The crystallization is absent in the exact ground state, but can be found in the first excited doubly degenerate level, whose energy converges to the ground state energy in the thermodynamic limit. Within this first excited energy level, we have constructed, explicitly, a many-body vector which displays exactly the same particle density as the mean field ground state, therefore establishing a precise connection between the mean field and exact solutions.

We have also demonstrated that the model exhibits a band insulator-metal transition. An intriguing question arises. Are we witnessing the transition from a band insulator into the Luttinger liquid? According to the general accepted view, the Luttinger liquid state is relevant to all un-gapped 1D fermions, so one would be inclined to answer positively to this question. But the model presented here is in many respects complementary to the Luttinger model, therefore we are inclined to believe that the metallic state is rather coupled to the charge-density-wave state. This issue will be investigated in the near future.

At the end, we should mention that there are hopes that the exactly solvable model  is stable enough to allow perturbative extensions that touch on realistic systems. For example, replacing the two perfectly flat bands with weakly dispersive bands should not pose major problems. Small variations of $v_k$ seem also to pose no major difficulties, but it is not clear at this point if such variations can be pushed to the limit of realistic interactions.  Clarifying these points will be important for understanding the impact of the model on the phase diagram of the 1D fermionic systems.   \medskip

{\bf References}\medskip

\bibliographystyle{iopart-num}

\providecommand{\newblock}{}

\end{document}